\documentclass[12pt]{article}
\usepackage{epsf,afterpage,hhline,array}
\title{\bf Hybrid adiabatic potentials in the QCD string model}
\author{
Yu.S.Kalashnikova \thanks{e-mail: yulia@heron.itep.ru\qquad
$^{**}$e-mail: kuzmenko@heron.itep.ru}, D.S.Kuzmenko$^{**}$}
\date{\it Institute of Theoretical and Experimental
Physics,\\ 117218, B. Cheremushkinskaya 25, Moscow, Russia}
\newcommand{\be} {\begin{equation}}
\newcommand{\ee} {\end{equation}}
\newcommand{\bdm} {\begin{displaymath}}
\newcommand{\edm} {\end{displaymath}}
\newcommand{\bc} {\begin{center}}
\newcommand{\ec} {\end{center}}
\newcommand{\beqa} {\begin{eqnarray}}
\newcommand{\eeqa} {\end{eqnarray}}

\newcommand{\lb} {\label}
\newcommand{\bfig} {\begin{figure}}
\newcommand{\btab} {\begin{tabular}}
\newcommand{\etab} {\end{tabular}}

\newcommand{\veR}{\mbox{\boldmath${\rm R}$}}
\newcommand{\ver}{\mbox{\boldmath${\rm r}$}}
\newcommand{\vep}{\mbox{\boldmath${\rm p}$}}
\newcommand{\vew}{\mbox{\boldmath${\rm w}$}}
\newcommand{\veL}{\mbox{\boldmath${\rm L}$}}
\newcommand{\vez}{\mbox{\boldmath${\rm z}$}}
\newcommand{\vej}{\mbox{\boldmath${\rm j}$}}
\newcommand{\veS}{\mbox{\boldmath${\rm S}$}}
\newcommand{\veB}{\mbox{\boldmath${\rm B}$}}
\newcommand{\e}{\mathrm{e}}
\newcommand{\mr}[1]{\mathrm{#1}}
\newcommand{\erf}{\mathrm{erf}(x)}
\newcommand{\ep}{\frac{\e^{-x^2}}{\sqrt{\pi}}}
\newcommand{\epa}[3]{\frac{#1\,\e^{-x^2}}{#2\sqrt{\pi}\,#3}}
\newcommand{\epx}{\frac{\e^{-x^2}}{\sqrt{\pi}\,x}}
\newcommand{\xep}{\frac{x\,\e^{-x^2}}{\sqrt{\pi}}}
\newcommand{\fr}[2]{\frac{#1}{#2}}
\newcommand{\trf}{\left<\tilde r_1\right>}
\newcommand{\trg}{\left<\frac{1}{\tilde r_1}\right>}
\newcommand{\lt}{\left}
\newcommand{\rt}{\right}
\newcommand{\LS}{\mathrm{LS}}
\newcommand{\stc}{\mathrm{sc}}
\newcommand{\VLS}[1]{V^{\LS\,(\mr{np})}_{#1}}
\newcommand{\VLSp}[1]{V^{\LS\,(\mr{p})}_{#1}}
\begin{document}
\maketitle

\begin{abstract}
The short- and intermediate-distance behaviour of the hybrid
adiabatic potentials is calculated in the framework of the QCD string
model. The calculations are performed with the inclusion of Coulomb
force. Spin-dependent force and the so-called string correction term
are treated as perturbation at the leading potential-type regime.
Reasonably good agreement with lattice measurements takes place for
adiabatic curves excited with magnetic components of field strength
correlators.
\end{abstract}

\section{Introduction}

Gluonic degrees of freedom in the nonperturbative region should manifest
themselves as QCD bound states containing constituent glue, so one expects 
that purely gluonic hadrons (glueballs) should exist as well as hybrids
where the glue is excited in the presence of $q\bar q$ pair.
General agreement is that the lightest hybrids occur in the mass
range between 1.3 and 1.9 GeV, so the absolute mass scale remains somewhat
imprecise in the absence of exact analytic methods of nonperturbative QCD.
Existing experimental data seem to point towards gluonic excitations being 
present, and {\it prima facie} candidates are identified \cite{data}, but
the conclusive evidences have never been presented. There is no hope that
in the nearest future data analyses could unambigiously pin-point
the signatures for gluonic mesons and settle the issue of 
constituent glue. The state-of-art is that the predictions of 
different models on hadronic spectra and decays are involved in order to
distinguish between gluonic mesons and conventional ones.

In such situation the lattice gauge calculations remain the only
source of knowledge. Lattice calculations are now accurate enough to
serve as a guide, so that the results of different QCD-motivated 
approaches can be compared and contrasted with lattice data. Of particular 
interest are the measurements of gluelump \cite{gluelump} and hybrid 
adiabatic potentials \cite{lattice}. These simulations measure the 
spectrum of the glue in the presence of static source in the adjoint 
colour representation (gluelump) and in the presence of static quark and 
antiquark separated by some distance $R$. These systems are the simplest 
ones and play the role of hydrogen atom of soft glue studies, as, first, 
the gluonic effects are not obscured by light dynamical quarks, and, 
second, the problem of centre-of-mass motion separation is not relevant 
here.

The large distance limit of hybrid adiabatic potentials is important, as 
one expects the formation of confining string at large $R$. The short 
range limit is relevant to the heavy hybrid mass estimations. One expects 
that in the case of very heavy quarks the hybrid resides in the bottom of 
the potential well given by the adiabatic curve, which, in accordance with 
lattice results \cite{lattice}, is somewhere around 0.25 fm for lowest 
curves. 

In the present paper we study hybrid adiabatic potentials in the so-called 
QCD string model \cite{lisbon}. This model deals with quarks 
and point-like gluons propagating in the confining QCD vacuum, and is 
derived from the Vacuum Background Correlators method. In the latter, 
the confining vacuum is parametrised by the set of gauge-invariant field 
strength correlators \cite{review} responsible, among other phenomena, for 
the area law asymptotics. The basic assumption of the QCD string model is 
the minimal area law for the Wilson loop, so that the only nonperturbative 
input is the string tension $\sigma$. 

The QCD string model describes the spectra of $q \bar q$ mesons with 
remarkable agreement \cite{mesons1},\cite{mesons2}. It is also quite 
successfull in 
describing glueballs \cite{glueball}, hybrids \cite{hybrid} and
gluelump \cite{Sgluelump}, as well as meson-hybrid-glueball mixing 
\cite{mixing}.

First studies of hybrid adiabatic potentials in the QCD string model
were performed in \cite{vibr}, with special attention paid to the 
large-distance limit. It was shown that at large interquark 
distances two kinds of QCD string vibrations take place, the 
potential-type longitudinal and string-type transverse ones. Here we consider 
the short-distance behaviour of the excitation curves.

The paper is organised as follows. In Section 2 we briefly discuss the 
essentials of the QCD string approach for gluons. The effective
Hamiltonian for a gluon bound by the static quark-antiquark pair
is derived in Section 3. It is argued in Section 4 that at short and
intermediate interquark distances the potential-type regime of
string vibrations is adequate, and the lowest excitation curves are
calculated. The spin-dependent forces and the so-called string corrections 
are considered in Section 5.  Results and discussion are given in
Section~6 together with conclusions and outlook.
Appendices contain the details of our variational calculations.

\section{Gluons in the confining background}       

The QCD string model for gluons is derived in the framework of 
perturbation theory in the nonperturbative confining background
\cite{pert}. The main idea is to split the gauge field as 
\be
A_\mu = B_\mu + a_\mu,
\lb{split}
\ee 
which allows to distinguish clearly between confining field configurations 
$B_\mu$ and confined valence gluons $a_\mu$. The valence gluons are 
treated as perturbation at the confining background.

We start with the Green function for the gluon propagating in the given
external field $B_\mu$ \cite{pert}:
\be
G_{\mu\nu} (x,y) = (D^2(B)\delta_{\mu\nu}+2ig F_{\mu\nu} (B))^{-1},
\lb{gf}
\ee
where both covariant derivative $D_{\lambda}^{ca}$ and field strength 
$F_{\mu\nu}^a$ depend only on the field $B_\mu$:
\be
D_{\lambda}^{ca}(B) = \delta^{ca} \partial_{\lambda} + g f^{cba} 
B^b_{\lambda},
\lb{B}
\ee
\be
F_{\mu\nu}^a(B) = \partial_{\mu} B^a_{\nu} - \partial_{\nu} B^a_{\mu} + g
f^{abc} B^b_{\mu} B^c_{\nu}.
\lb{F}
\ee
The term, proportional to $F_{\mu\nu}(B)$ in (\ref{gf}) is responsible for 
the gluon spin interaction. We neglect it for a moment, it will be 
considered in Section 5.

Now we use Feynman-Schwinger representation for the quark-antiquark-gluon
Green function \cite{hybrid}, which, for static quark and antiquark, is 
reduced to the form
\be
G(x_g, y_g)= \int^\infty_0 ds \int Dz_g \exp (-K_g) \langle
{\cal W}\rangle_B,
\lb{FS}
\ee
where angular brackets mean averaging over background field. The quantity 
$K_g$ is the kinetic energy of gluon (to be specified below), and all the 
dependence on the vacuum background field is contained in the generalized 
Wilson loop $\cal W$, depicted in Fig.1, where the contours $\Gamma_Q$ and 
$\Gamma_{\bar Q}$ run over the classical trajectories of static quark and 
antiquark, and the contour $\Gamma_g$ runs over the gluon trajectory
$z_g$ in (\ref{FS}).
\begin{figure}[!t]
\epsfxsize=8cm
\hspace*{4.35cm}
\epsfbox{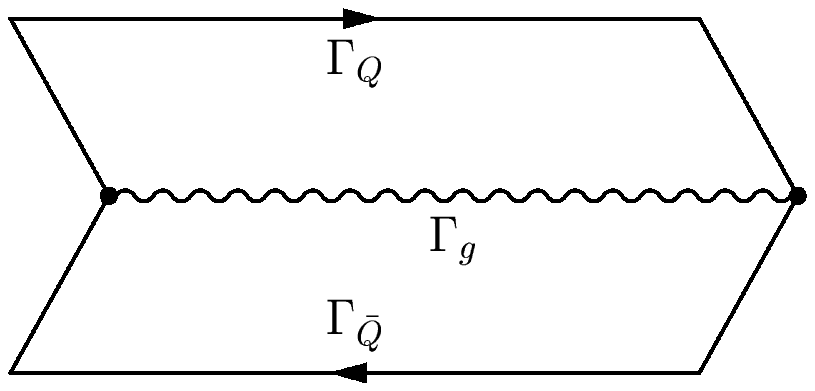}
\caption{Hybrid Wilson loop}
\label{hWloop}
\end{figure}

The expression (\ref{FS}) is the starting point of the QCD string model,
as, under the minimal area law assumption, the Wilson loop configuration 
takes the form
\be
\langle {\cal W}\rangle_B = \frac{N_c^2-1}{2}\exp (-\sigma(S_1+S_2)),
\lb{area}   
\ee
where $S_1$ and $S_2$ are the minimal areas inside the contours formed by 
quark and gluon and antiquark and gluon trajectories correspondingly.

\section{Einbein field form of the gluonic Hamiltonian}

To proceed further we are to fix the gauge in the reparametrization 
transformations group. For the case of static quark and antiquark sources 
the most natural way to do this is to identify the proper time $\tau$
of the Feynman-Schwinger representation with the laboratory time. Then the 
classical quark and antiquark trajectories are given by
\be
z_{Q\mu}=(\tau, \frac{\veR}{2})~~,~~
z_{\bar Q \mu}=(\tau, -\frac{\veR}{2})~~,
\lb{Q}
\ee
and the action of the system can be immediately obtained from the 
representation (\ref{FS}):
$$
A=\int^T_0 d\tau \left \{-\frac{\mu}{2}+\frac{\mu\dot r^2}{2}- \sigma
\int^1_0 d\beta_1
\sqrt{(\dot w_1 w_1')^2-\dot w_1^2w^{'2}_1}-\right.
$$
\be
\left.-\sigma\int^1_0 d\beta_2
\sqrt{(\dot w_2 w_2')^2-\dot w_2^2w^{'2}_2}\right \}.
\lb{action}
\ee
Here $\ver$ is the three-dimensional gluonic coordinate, and the minimal 
surfaces $S_1$ and $S_2$ are parametrized by the coordinates
$w_{i\mu}(\tau,\beta_i) ,~ i=1,2,~ \dot w_{i\mu}= \frac{\partial w_{i\mu}}
{\partial \tau},~ w'_{i\mu}= \frac{\partial w_{i\mu}}{\partial \beta_i}.$
Choosing the straight-line ansatz for the minimal surfaces one 
has in the laboratory time gauge
\be
w_{i0}=\tau, ~~~\vew_{1,2} = \pm (1-\beta) \frac{\veR}{2}+\beta
\ver.
\lb{w}
\ee

The kinetic energy in (\ref{action}) is given in the so-called einbein 
field form \cite{einbein}. The einbein field $\mu=\mu(\tau)$ is the 
auxiliary field introduced to deal with relativistic kinematics. Note
that in the case of gluon one is forced to introduce it from the very
beginning, as it provides the meaningful dynamics for the massless 
particle.

In order to pass to the Hamiltonian formulation it is convenient to
get rid of Nambu-Goto square roots in (\ref{action}), introducing 
continuous set of einbein fields $\nu_i=\nu_i(\tau,\beta_i)$, as it was 
first suggested in \cite{mesons1}:
$$
L=-\frac{\mu}{2}+\frac{\mu\dot r^2}{2}-
\int^1_0 d\beta_1\frac{\sigma^2 r^2_1}{2\nu_1}-\int^1_0
d\beta_1\frac{\nu_1}{2}(1-\beta_1^2 l_1^2)-
$$
$$
-\int^1_0 d\beta_2\frac{\sigma^2 r^2_2}{2\nu_2}-\int^1_0
d\beta_2\frac{\nu_2}{2}(1-\beta_2 l_2^2),
$$
\be
l^2_{1,2}=\dot r^2-\frac{1}{r^2_{1,2}}(\ver_{1,2}\dot{\ver})^2,~~
\ver_{1,2}=\ver\pm \frac{\veR}{2}.
\lb{lagr}
\ee

Note, that the Lagrangian (\ref{lagr}) describes the constrained system.
As no time derivatives of the einbeins enter it, the corresponding 
equations of motion play the role of second-class constraints 
\cite{einbein}.

The Hamiltonian $H=\vep\dot{\ver}-L$ is easily obtained from the 
Lagrangian (\ref{lagr}):
\be
H=H_0+\frac{\mu}{2}
+\int^1_0 d\beta_1\frac{\sigma^2 r^2_1}{\nu_1}
+\int^1_0 d\beta_2\frac{\sigma^2 r^2_2}{\nu_2}
+\int^1_0 d\beta_1\frac{\nu_1}{2}+\int^1_0 d\beta_2\frac{\nu_2}{2},
\lb{ham}
\ee

$$
H_0=\frac{p^2}{2(\mu+J_1+J_2)}+
$$
$$
\frac{1}{2\Delta (\mu+J_1+J_2)}
\left\{ \frac{(\vep\ver_1)^2}{r^2_1}J_1(\mu+J_1)+
\frac{(\vep\ver_2)^2}{r^2_2}J_2(\mu+J_2)+\right.
$$
\be
\left.\frac{2J_1J_2}{r^2_1r^2_2}(\ver_1\ver_2)(\vep\ver_1)(\vep\ver_2)
\right\}
\lb{ham0}
\ee

$$
\Delta=(\mu+J_1)(\mu+J_2)-
J_1J_2\frac{(\ver_1\ver_2)^2}{r^2_1r^2_2},~~
J_i=\int^1_0 d\beta_i\beta_i^2\nu_i(\beta_i),~~ i=1,2.
$$

The Hamiltonian (\ref{ham}) together with the constraints
\be
\frac{\partial H}{\partial\mu}=0,~~
\frac{\delta H}{\delta \nu_i(\beta_i)}=0
\lb{constraints}
\ee
completely defines the dynamics of the system at the classical level.
To quantize one should first find the extrema of einbeins from the 
equations (\ref{constraints}) and substitute them back to the Hamiltonian 
(\ref{ham}). Then, the extremal values of einbeins would become the 
nonlinear operator functions of coordinate and momentum, and, in addition, 
the problem of operator ordering would arise. To avoid this complicated
problem the approximate einbein field method is usually applied in the
QCD string model calculations. Namely, einbeins are treated as $c$ number
variational parameters: the eigenvalues of the Hamiltonian (\ref{ham}) are
found as functions of $\mu$ and $\nu_i$, and minimized with respect to 
einbeins to obtain the physical spectrum. Such procedure, first suggested 
in \cite{mesons1}, provides the accuracy of about 5-10\% for the groud
state (for the details see first entry in \cite{mesons2}).

\section{Potential regime of the QCD string vibrations}

The einbeins $\mu$ and $\nu_i(\beta_i)$ play the role of constituent 
gluon mass and energy densities along two strings. Note that even with 
simplifying assumptions of the einbein field method these quantities
are not introduced as model parameters, but are calculated in the
formalism. It is clear from the form (\ref{ham0}) of the kinetic
energy that two kinds of motion compete to form the spectrum: the
potential-type longitudinal with respect to $\veR$ vibrations due to 
gluonic mass $\mu$ and  string-type transverse ones due to the string 
inertia. 

It was shown in \cite{vibr} that for large interquark distances, $R\gg
1/\sqrt{\sigma}$, these two types of motion decouple, displaying the
corrections to the leading $\sigma R$ behaviour proportional to $(\frac
{\sigma}{R})^{1/3}$ in the case of longitudinal vibrations and 
proportional to $\frac{1}{R}$ for transverse ones. 

On the contrary, for small $R$ one can neglect the terms $J_i$
responsible for string inertia in the kinetic energy (\ref{ham0}).
Then the Hamiltonian takes the form
\be
H=\frac{p^2}{2\mu}+\frac{\mu}{2}+
\int^1_0d\beta_1\frac{\sigma^2r^2_1}{2\nu_1}+
\int^1_0d\beta_2\frac{\sigma^2r^2_2}{2\nu_2}+    
\int^1_0d\beta_1\frac{\nu_1}{2}+
\int^1_0d\beta_2\frac{\nu_2}{2}.
\lb{hame}
\ee
The spectrum of the Hamiltonian (\ref{hame}) was found in
\cite{vibr}. At small $R$ it reads
\be
E_n(R)=2^{3/2}\sigma^{1/2}
\lt(n+\frac32\rt)^{1/2}+
\frac{\sigma^{3/2}R^2}{2^{3/2}(n+\frac32)^{1/2}}.
\lb{Ee}
\ee
The extremal values of einbeins are given by
\be
\mu_n(R)=2^{1/2}\sigma^{1/2}
\lt(n+\frac32\rt)^{1/2}-\frac{\sigma^{3/2}R^2}
{2^{5/2}(n+\frac32)^{1/2}},
\lb{mue}
\ee
\be
\nu_{1,2 n}(R)=
\frac{(n+\frac32)^{1/2}\sigma^{1/2}}
{2^{1/2}}+\frac{3\sigma^{3/2}R^2}{2^{7/2}
(n+\frac32)^{1/2}},
\lb{nue}
\ee
where 
$n$ is the number of oscillator quanta. 

\begin{figure}[!t]
\epsfxsize=14cm
\hspace*{1.35cm}
\epsfbox{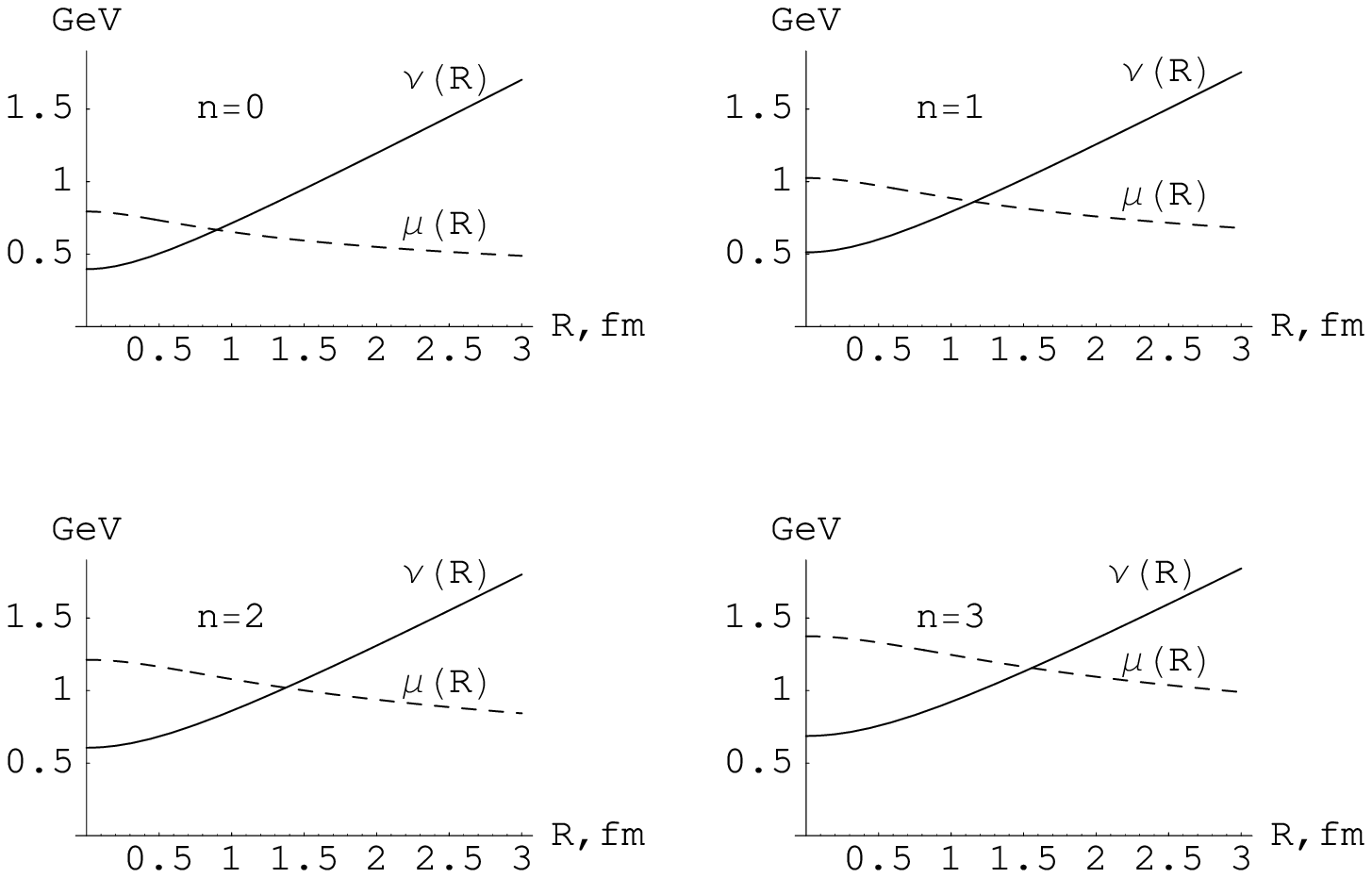}
\caption{Einbein fields  $\nu_n(R)$ (solid curve)  and
$\mu_n(R)$ (dashed curve) for  $n=0,1,2,3$ and
$\sigma=0.21$ GeV$^2$. }
\label{munu}
\end{figure}

The expressions (\ref{mue}), (\ref{nue}) immediately yield $J_{1,2}/\mu
\approx \frac16$, so the neglect of string inertia is justified. The 
 curves $\mu_n(R)$ and $\nu_n(R)$ for arbitrary $R$ from \cite{vibr}
are shown at Fig.\ref{munu} for $n=0,1,2,3$ and $\sigma = 0.21$ GeV.
It is clear that the potential-type Hamiltonian can be employed at $R
\leq 1$ fm for $n=0,1$ and at $R \leq 1.5$ fm for $n=2,3$, and the
corrections due to string inertia can be taken into account
perturbatively.

The form (\ref{hame}) allows to eliminate einbeins and arrive at the
potential-type Hamiltonian
\be
H=\sqrt{p^2}+\sigma r_1+\sigma r_2.
\lb{potham}
\ee
Nevertheless, as we are going to calculate the spin-dependent forces and 
string correction, we prefer to eliminate only einbeins $\nu_i$, treating
the quantity $\mu$ in the framework of the einbein field method.  

If only confining force is taken into account, the QCD string model
predicts the oscillator potential (\ref{Ee}) with the minimum at $R=0$. 
However, the minimum is shifted, if the long-range confining force is 
augmented by the short-range Coulomb potential, 
\be
V_c=-\frac{3\alpha_s}{2r_1}-\frac{3\alpha_s}{2r_2}+\frac{\alpha_s}{6R}.
\lb{coul}
\ee
The coefficients in (\ref{coul}) are in accordance with the colour content
of the $Q\bar Q g$ system \cite{H&M}. The $Q\bar Q$ Coulomb force 
in (\ref{coul}) is repulsive, and it is compatible with the behaviour of 
gluon energies \cite{lattice} at small $R$. Note, that such behaviour
comes out naturally in the QCD string model, as point-like gluon does
carry colour quantum numbers.

The final form of our Hamiltonian reads
\be
H=\frac{p^2}{2\mu}+\frac{\mu}{2}+\sigma r_1+\sigma r_2+V_c.
\lb{hamf}
\ee
The angular momentum is not conserved in the Hamiltonian (\ref{hamf}),
but it is a good quantum number in the einbein field Hamiltonian
(\ref{hame}). For the case of pure confining force we have compared the 
spectra of exact and einbein-field Hamiltonians, and have found that 
angular momentum is conserved within better than 5\% accuracy. The same 
phenomenon is observed in the constituent gluon model \cite{swanson},
and seems to be a consequence of linear potential confinement embedded 
there.

The eigenvalue problem for the Hamiltonian (\ref{hamf}) was solved 
variationally with wave functions 
\be
\vec\Psi_{jl\Lambda}(\ver)=\phi_l(r)\sum_{\mu_1\mu_2}
C^{j\Lambda}_{l\mu_1 1\mu_2} Y_{l\mu_1}(\frac{\ver}{r})
\vec\chi_{1\mu_2},
\lb{Psi}
\ee
where $\vec\chi_{1\mu}$ is the spin 1 wave function,
$\Lambda=\left\vert\frac{\vej\veR}{R}\right\vert$ is the projection of 
total angular momentum $\vej$ onto $z$ axis, $\vez\parallel\veR$. The
radial wave functions $\phi_l(r)$ were taken to be Gaussian, that is of 
the form $\exp(-\beta^2 r^2/2)$ multiplied by the appropriate
polynomials, with $\beta$ treated as variational parameter. The 
eigenvalues $E_{jl\Lambda}(\mu,R)\equiv
\langle \vec\Psi_{jl\Lambda}|H(\mu,R)|\vec\Psi_{jl\Lambda}\rangle$
were found in such a way, and the resulting adiabatic potentials,
\be
V^0_{jl\Lambda}(R)=E_{jl\Lambda}(\mu^{*}(R),R),
\lb{E}
\ee
depend on the extremal value $\mu^{*}$ defined from the condition
\be
\frac{\partial E_{jl\Lambda}(\mu,R)}{\partial \mu}=0. 
\lb{mu*}
\ee
The details
of this variational procedure are given in the Appendix A.

In the QCD string model the gluon is effectively massive, and has three 
polarizations \cite{glueball},\cite{Sgluelump}. Only two of them are 
excited with magnetic components of field strength correlators, 
used in lattice calculations. We list these states in Table 1, in terms 
of $j,l,\Lambda$ and standard notations borrowed from physics of diatomic 
molecules. For more details justifying such correspondence see 
\cite{Sgluelump}.

\begin{table}[t!]
\centering
\small
\caption{Quantum numbers of lowest levels}
\vspace{0.3cm}
\label{table}
\begin{tabular}{||c|c|c|c||}
\hhline{|t:=:=:=:=:t|}
(a)\rule{0pt}{5mm} & $j^{PC}=1^{+-}$ & $j=1,~l=1,~\Lambda=0,1$ &
$\Sigma_u^-,~\Pi_u$ \\
(b) & $j^{PC}=1^{--}$ &
$j=1,~l=2,~\Lambda=0,1$ & $\Sigma_g^+,~\Pi_g$ \\
(c) &
$j^{PC}=2^{--}$ & $j=2,~l=2,~\Lambda=0,1,2$ &
$\Sigma_g^-,~\Pi_g,~\Delta_g$ \\
(d) & $j^{PC}=2^{+-}$ &
$j=2,~l=3,~\Lambda=0,1,2$ & $\Sigma_u^+,~\Pi_u,~\Delta_u$ \\[5pt]
\hhline{|b:=:=:=:=:b|}
  \end{tabular}
\label{qnum}
\end{table}

Fitting the ground lattice state 
with Coulomb+linear potential yields the values of parameters
 $\alpha_s=0.4$ and $\sigma=0.21$ GeV$^2$.

\section{String correction and spin-dependent interaction}

Now we turn to the calculations of corrections to the leading potential
regime (\ref{E}). 

Let us first consider the correction due to string inertia. It corresponds 
to the terms in (\ref{ham0}) linear in $J_i$:
\be
H^{\stc} = -\frac{J_1+J_2}{2\mu^2}p^2+\frac{1}{2\mu^2}
\left( \frac{(\vep\ver_1)^2}{r_1^2}J_1+\frac{(\vep\ver_2)^2}{r_2^2}J_2
\right).
\lb{stJ}
\ee
In the potential regime $\nu_i=\sigma r_i$, and $J_i=\sigma r_i/3$. So the
string correction Hamiltonian takes the form
\be
H^{\stc}=-\frac{\sigma}{6\mu^2}\left( \frac{1}{r_1^2}L_1^2 +
\frac{1}{r_2^2} L_2^2 \right),
\lb{stf}
\ee
where
\be
\veL_i=\ver_i \times \vep.
\lb{Li}
\ee
The choice (\ref{Li}) solves the ordering problem in (\ref{stf}), as it 
assures the hermiticity of the operator $H^{\stc}$.
 
In actual calculations it is convenient to rewrite (\ref{stf}) as
\be
H^{\stc}=-\frac{\sigma}{6\mu^2}\left\{ \left(
\frac{1}{r_1^2}+\frac{1}{r_2^2} \right) \left(L^2+\frac{R^2}{4}H_1 \right)
+\left( \frac{1}{r_1^2}-\frac{1}{r_2^2} \right) RH_2 \right\},
\lb{H1H2}
\ee
where
\be
H_1=- \left( \partial^2_\rho + \frac{1}{\rho}\,\partial_\rho +
\frac{1}{\rho^2}\,\partial^2_\phi \right),
\lb{h1}
\ee
\be
H_2=zH_1+\rho\,\partial_\rho\partial_z+\partial_z.
\lb{h2}
\ee
The adiabatic potentials of string correction are listed in the
Appendix B.

The spin-dependent force
originates from the term, proportional to $F_{\mu\nu}(B)$ in (\ref{gf}).
This term generates the spin-dependent interaction, as
\be
-iF_{ik}=(\veS \veB)_{ik},
\lb{spin}
\ee 
where spin operator $\veS$ acts at the vector function $\vec \Psi$ as
\be
\left(S_i \vec\Psi\right)_j = -i\epsilon_{ijk}\Psi_k.
\lb{S}
\ee
One averages this term over the background, as it was done  in
\cite{spin}, and, in our case, it gives
\be
H^{\LS\,\mr{(np)}}= -\frac{\sigma}{2\mu^2} \left( \frac{1}{r_1}(\veL_1
\veS) +\frac{1}{r_2}(\veL_2 \veS) \right),
\lb{th1}
\ee 
\be
H^{\LS\,\mr{(p)}}= \frac{3\alpha_s}{4\mu^2} \left(
\frac{1}{r_1^3}(\veL_1 \veS) +\frac{1}{r_2^3}(\veL_2 \veS) \right),
\lb{th2}
\ee
for nonperturbative and perturbative forces respectively. One
easily recognizes the contribution of Thomas precession in
(\ref{th1}), (\ref{th2}).
\begin{figure}[!h]
\epsfxsize=12cm
\hspace*{2.35cm}
\epsfbox{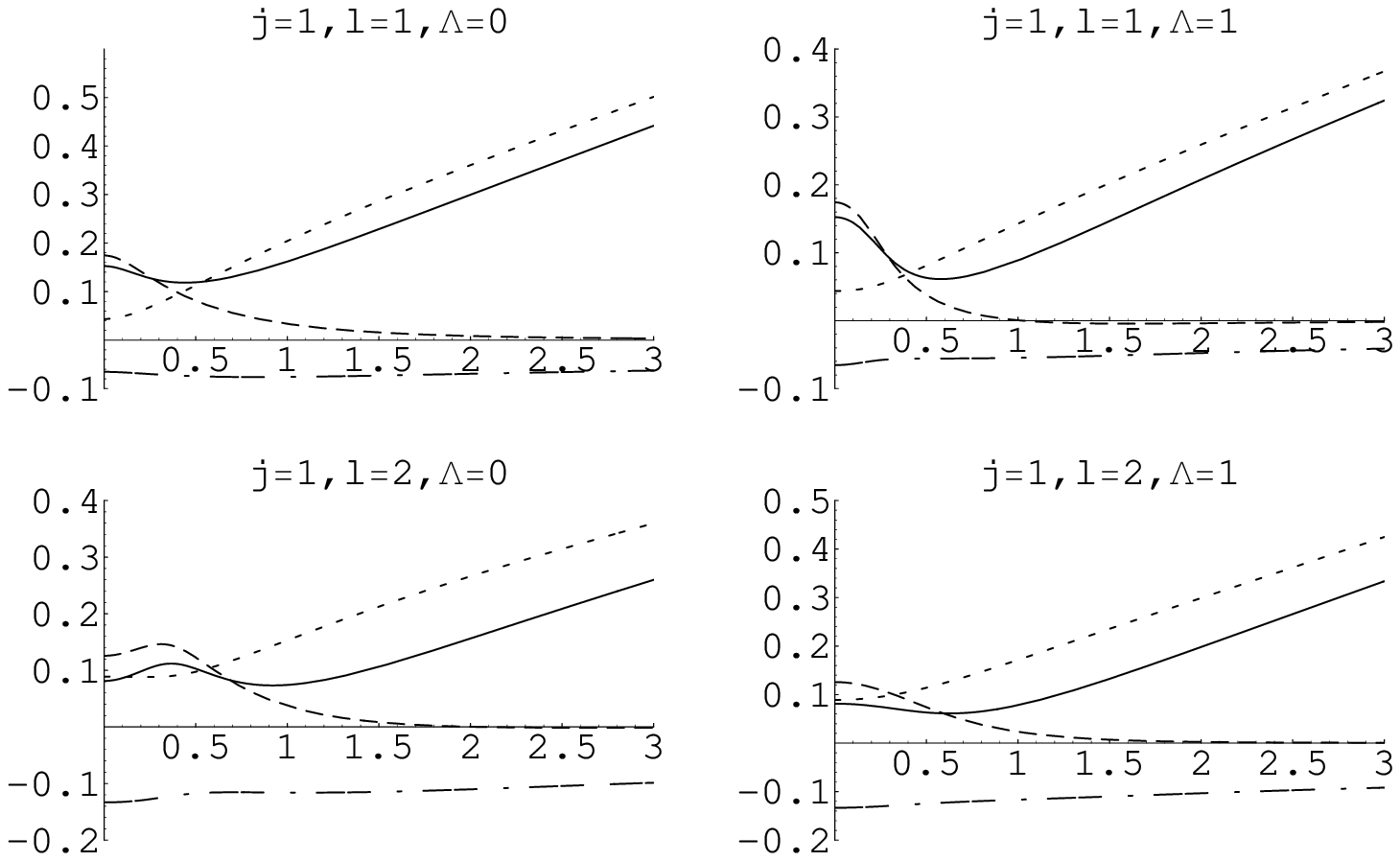}
\caption{Taken with opposite sign potentials of string $-V^{\stc} (R)$
(dotted curves), spin-dependent perturbative $-\VLSp{}(R)$
(dashed curves) and nonperturbative  $-\VLS{}(R)$ (dashed-dotted
curves) corrections  along with their sum $-V^{\stc}
(R)-\VLSp{}(R)-\VLS{}(R)$ (solid curves), in GeV, for levels (a) and
(b) of Table 1. $R$ is measured in fm.}
\label{cor1}
\end{figure}
\begin{figure}[!h]
\epsfxsize=15cm
\hspace*{0.85cm}
\epsfbox{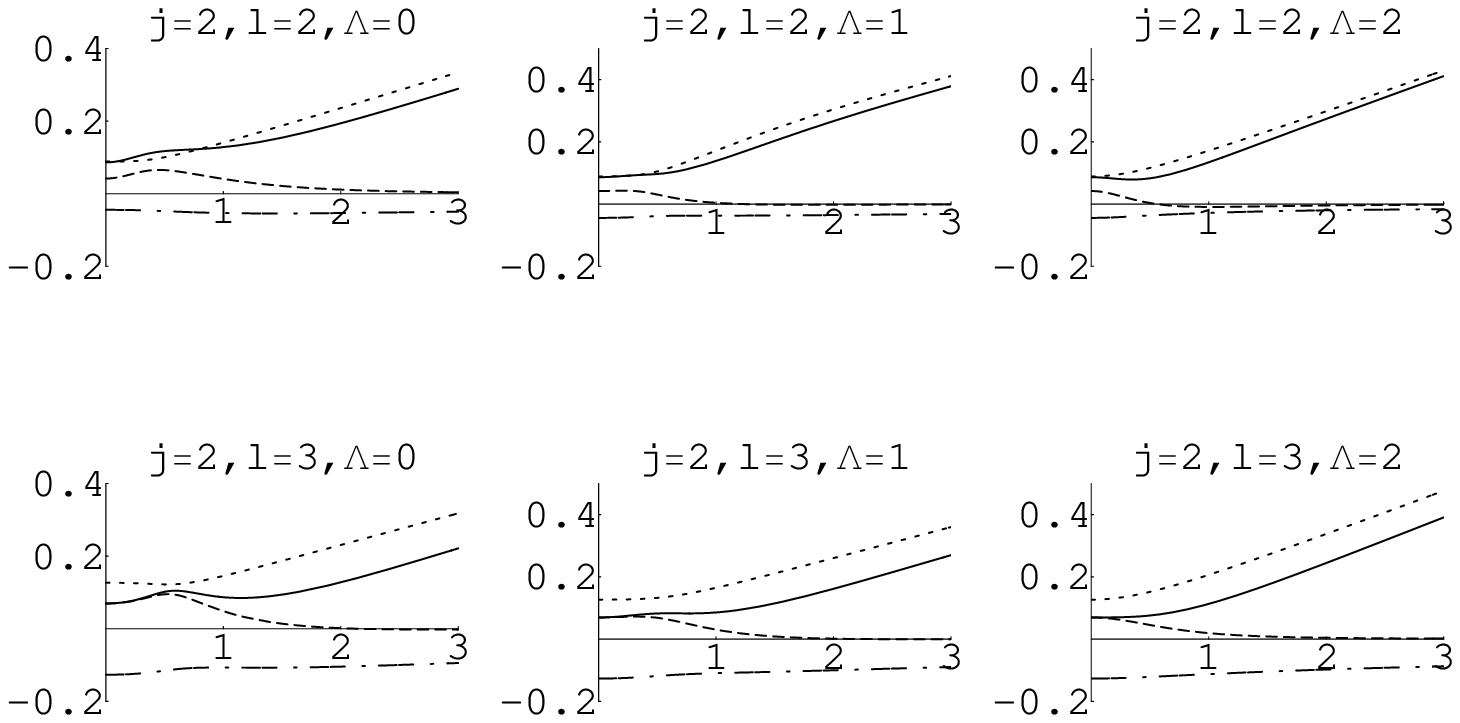}
\caption{The same as in Fig. \ref{cor1}, for levels (c) and (d).}
\label{cor2}
\end{figure}
The spin-dependent interaction is conveniently represented as
\be
H^{\LS\,\mr{(np)}}= -\frac{\sigma}{2\mu^2} \left\{ \left(
\frac{1}{r_1}+\frac{1}{r_2} \right) (\veL\veS) + \left( \frac{1}{r_1}
- \frac{1}{r_2} \right) \frac R2 H_R \right\},
\lb{LSnp}
\ee
\be
H^{\LS\,\mr{(p)}}= \frac{3\alpha_s}{4\mu^2} \left\{ \left(
\frac{1}{r_1^3}+\frac{1}{r_2^3} \right) (\veL\veS) + \left(
\frac{1}{r_1^3}-\frac{1}{r_2^3} \right) \frac R2 H_R \right\},
\lb{LSp}
\ee
where
\be
H_R= \frac{\e^{-i\phi}}2(-\partial_\rho+\frac{i}{\rho}\,
\partial_\phi)S_{+}+\frac{\e^{i\phi}}2(\partial_\rho+\frac{i}{\rho}\,
\partial_\phi)S_{-},
\lb{HR}
\ee
$$
S_{\pm}=S_x \pm iS_y.
$$

Spin-dependent potentials are given in the Appendix C.

 We would like to stress here, that, in spite of apparently
nonrelativistic form of the expressions (\ref{H1H2}) and (\ref{LSnp}),
(\ref{LSp}),
these are not the nonrelativistic inverse mass expansion. The mass
$\mu$ entering these expressions is replaced in matrix elements by the
value $\mu^{*}$, obtained from stationary point equation (\ref{mu*}).
The latter plays the role of effective mass of the gluon, and is not
large. The $R$-dependence of corrections is shown at Figs.\ref{cor1},
\ref{cor2}.

\section{Results and discussion}

\begin{figure}[!t]
\epsfxsize=15cm
\hspace*{0.85cm}
\epsfbox{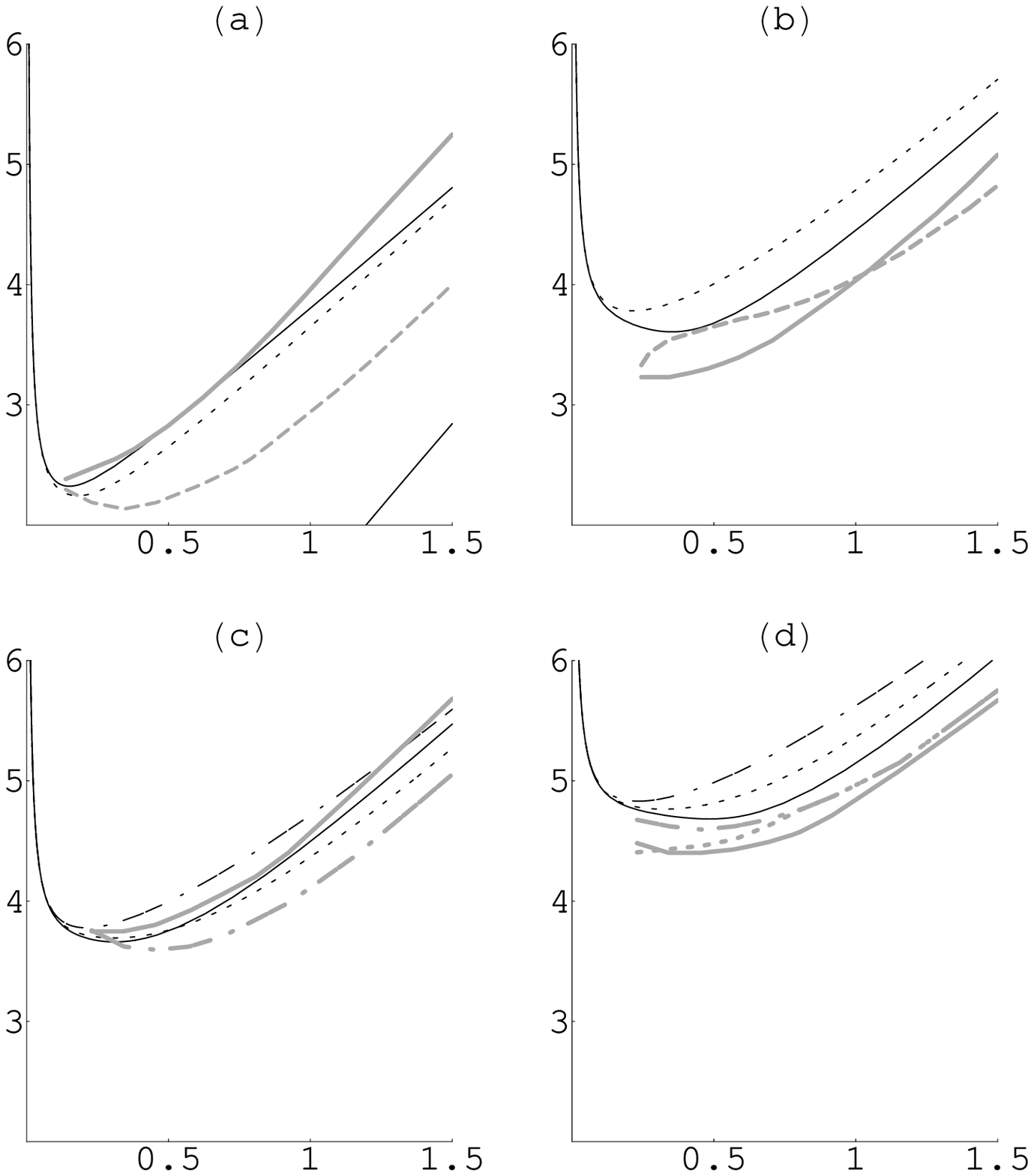}
\caption{Hybrid potentials with corrections included (black curves)
compared to lattice ones (thick grey curves)
in units $1/r_0=$ 400 MeV ($r_0=2.5$ GeV$^{-1}$).  $Q \bar
Q$-distance $R$ is measured in $2r_0\approx$ 1 fm. States
(a)--(d) are given in Table \ref{qnum}. Solid, dashed and
dashed-dotted curves correspond to $\Lambda=1,2$ and 3. Solid line
at right bottom corner of Fig.\ref{fullcor}(a) represents the
Coulomb+linear potential with $\alpha_s=0.4$ and $\sigma=0.21$
GeV$^2$.}
\label{fullcor}
\end{figure}

The results of full QCD string calculations are given at
Fig.\ref{fullcor} for the excitation curves listed in Table 1,
together with the lattice data \cite{lattice}.
 As it is seen from
the Fig.\ref{fullcor}, at small distances the calculated curves are in
good agreement with lattice results, with the accuracy better that 100
MeV. The level ordering is reproduced too, with the exception of
$\Sigma^{-}_g$, $\Pi_g$ and $\Delta_g$ levels (Fig.\ref{fullcor}c).
Note, that the lattice data claim only one $\Pi_g$ level, and its 
$R$-behaviour is rather peculiar  (see dashed thick grey curve at
Fig.\ref{fullcor}b). One expects that the curves should tend to form degenerate
levels as the distance $R$ decreases, fulfilling the angular momentum
conservation demands at $R=0$. This feature is made explicit in the
QCD string model: our calculations reproduce the gluelump spectrum
\cite{Sgluelump} for $R=0$ after subtracting the $Q\bar Q$ Coulomb
force (last term in (\ref{coul})) and with obvious replacement $2\sigma
\rightarrow \sigma_{adj}=\frac{9}{4}\sigma$, $2\alpha_s \rightarrow
\frac{9}{4}\alpha_s$.   

 Lattice data indeed seem to follow such tendency. Moreover, the
curvatures of all potentials but $\Pi_g$ are compatible with the
dominance of Coulomb force acting between static sources in octet
colour representation. Thus one suspects \cite{nora}, \cite{bali},
that something goes wrong with the lattice $\Pi_g$ level, and its
strange behaviour could be due to the presence of several levels,
severely mixed and poorly resolved by present simulations.

 The most pronounced feature of the QCD string approach is the
following. It was already mentioned that the gluon here is effectively
massive, and has three polarizations. As a consequence, the level
ordering follows the increasing dimension of valence gluon operator,
or, in other words, the increasing orbital momentum $l$. This is in
contrast to the standard viewpoint of constituent glue studies, see
\cite{jaffe} and realization of this idea in the framework of
potential NRQCD \cite{nora}. The level ordering there is supposed to
follow the increasing dimension of the operators $E_i$, $B_i$,
$D_iB_k$ {\it etc}. The equations of motion, which relate different
operators with the same quantum numbers, are involved to exclude
spurios states in such picture.

In our approach, we expect an extra family of levels to appear, namely, 
$\Sigma^+_g$ and $\Pi_g$, and the corresponding gluelump limit achieved
with $1^{--}$ quantum numbers. The wave functions of these states contain 
mostly the $l=0$ component, so these levels should be the lowest one-gluon
ones. The search for this family, accessible only with electric field 
correlators, is of paramount importance both in gluelump and adiabatic 
potentials settings. The presence or absence of such states would allow to 
discriminate among models.

The flux-tube model \cite{fluxtube}, as well as its relativistic version
\cite{Olsson}, assumes that soft glue is string-like, with phonon-type 
effective degrees of freedom. These string phonons are colourless, so that 
the $Q\bar Q$ pair is in colour-singlet state. Thus the Coulomb $Q\bar Q$ 
interaction is to be attractive. Adiabatic potentials are calculated
in \cite{Olsson}, and, in order to improve the short-range 
behaviour of adiabatic curves, the Coulomb interquark repulsion was added, 
which obviously contradicts the general phylosophy of
flux-tube. Without essential modifications of dynamical picture at small 
interquark distances, the flux-tube-type models seem to be ruled out by 
lattice data \cite{lattice}.
A rather elaborated constituent gluon model \cite{swanson}, based on 
field-theoretical Hamiltonian approach, agrees with lattice data on hybrid
potentials at short and intermediate interquark distances only under 
rather confusing assumption of gluon parity taken to be positive. 

The gluelump spectrum, as well as small $R$-limit of hybrid adiabatic 
potentials, is successfully calculated in the bag model \cite{bag}.
The lowest bag-model gluelump state is $1^{+-}$, and the $\Sigma-\Pi$ 
splitting at small interquark distances is in accordance with lattice
data. In this regard we stress once more the importance of lattice 
measurements with electric field correlators. If the ground state 
$\Sigma^+_g-\Pi_g$ family is not found, then, with above-mentioned 
drawbacks of flux-tube and constituent-gluon pictures, it would 
mean that soft glue is bag-like rather than string-like or point-like. 

To conclude, we have presented full QCD string calculations 
of hybrid adiabatic potentials. The results are in general agreement with 
lattice data. We outline the problems, connected with restricted set of 
gluonic operators used in lattice simulations, and call for further
studies of excitation curves with electric field operators. 

We are grateful to Yu.A.Simonov for numerous stimulating discussions.
The financial support of RFFI grants 00-02-17836 and 00-15-96786,
and INTAS OPEN 2000-110, 2000-366 is acknowledged.

\appendix

\section{Variational calculations of adiabatic potentials}
\setcounter{equation}{0}
\renewcommand{\theequation}{A.\arabic{equation}}

In this Appendix we derive the explicit variational equations for
adiabatic potentials (\ref{E}).

The average momentum for oscillatory wave finction with orbital
momentum  $l$ is given as
\be
\left<p^2\right>_l=\frac{2l+3}{2}\beta^2,
\label{A1}
\ee
where $\beta$ is the variational parameter with dimension of mass.
Thus the calculation of extremum over $\mu$ (\ref{mu*}) leads to
expression
\be
\mu^*=\beta\sqrt{\frac{2l+3}{2}}.
\label{A2}
\ee
Due to the symmetry of wave functions,
\be
\left<r_1\right>=\left<r_2\right>,~~
\left<\frac{1}{r_1}\right>=\left<\frac{1}{r_2}\right>.
\label{A3}
\ee
Let us introduce the dimensionless variables
\be
x=\beta\, \frac{R}{2}~ \mbox{ É } ~\tilde r_1=\beta r_1
\label{A4}
\ee
and define the dimensionless averages
\be
\left<\tilde r_1\right>\equiv f(x),~~
\left<\frac{1}{\tilde r_1}\right>\equiv g(x).
\label{A5}
\ee
For the energy levels we will get
$$
 E(\beta,x)=\frac{1}{\beta}\, 2\sigma f(x)+\beta
\left(\sqrt{\frac{2l+3}{2}}- 3 \alpha_s g(x) +\frac{\alpha_s}{12
x}\right).
\label{A6}
$$
Let us perform now the extremum of the last equation over $\beta$
provided that $x$ is the function of $\beta$ and $\partial
f(x(\beta))/ \partial\beta= f'(x) x/\beta$. So we find that
\be
\beta^2(x)=\frac{2\sigma(f(x)-xf'(x))}{\displaystyle \sqrt{\frac{2l+3}{2}}-
3\alpha_s(g(x)+xg'(x))},
\label{A7}
\ee
\be
E(x)=\frac{2\sigma}{\beta(x)}f(x)+\beta(x)\left(\sqrt{\frac{2l+3}{2}}-
3 \alpha_s g(x) +\frac{\alpha_s}{12 x}\right),
\label{A8}
\ee
\be
R(x)=\frac{2x}{\beta(x)}.
\label{A9}
\ee
The equations (\ref{A8}), (\ref{A9}), along with calculated functions
$f(x),~g(x),~\beta(x)$ define explicitly adiabatic  potentials
(\ref{E}). Note that all expressions for the averages
contain the error function
\be
\erf=\fr2{\sqrt{\pi}}\int_0^x dt\,\e^{-t^2}.
\label{A10}
\ee
We present them below for the levels of Table \ref{qnum}.
\begin{eqnarray}
\rule{0pt}{10mm}\trf_{110}&=&x\lt\{\erf\lt(1+\fr1{x^2}-\fr1{4x^4}\rt)+
\epx \lt(1+\fr1{2x^2}\rt)\rt\}
\label{A11}
\nonumber\\
\trg_{110}&=&\fr1x\lt\{\erf\lt(1-\fr1{2x^2}\rt)+\epx\rt\}
\label{A12}
\\
\beta_{110}^2&=&\fr{\displaystyle \fr{4\sigma}x
\lt\{\erf\lt(1-\fr1{2x^2}\rt)+\epx\rt\}}
{\displaystyle \sqrt{\fr52}-\fr{3\alpha_s}{x^3}
\lt(\erf-\epa{2x}{}{}\rt)}
\label{A13}
\nonumber\\
\rule{0pt}{16mm}\trf_{111}&=&x\lt\{\erf\lt(1+\fr3{4x^2}+\fr1{8x^4}\rt)+
\epx \lt(1-\fr1{4x^2}\rt)\rt\}
\label{A14}
\nonumber\\
\trg_{111}&=&\fr1x\lt\{\erf\lt(1+\fr1{4x^2}\rt)
-\ep\lt(x+\fr1{2x}\rt)\rt\}
\label{A15}
\\
\beta_{111}^2&=&\fr{\displaystyle \fr{2\sigma}x
\lt\{\fr12\erf\lt(3+\fr1{x^2}\rt)-\ep\lt(x+\fr1x\rt)\rt\}}
{\displaystyle \sqrt{\fr52}-\fr{3\alpha_s}{x^3}\lt(-\fr12\erf+
\xep\lt(1+2x^2+2x^4\rt)\rt)}
\label{A16}
\nonumber\\
\rule{0pt}{16mm}\trf_{120}&=&x\lt\{\erf\lt(1+\fr{14}{15x^2}+\fr9{20x^4}\rt)+
\epx \lt(\fr{11}{15}-\fr9{10x^2}\rt)\rt\}
\nonumber\\
\trg_{120}&=&\fr1x\lt\{\erf\lt(1+\fr7{10x^2}\rt)
-\epa{x}{5}{}\lt(\fr{8x^2}3+\fr{28}3+\fr7{x^2}\rt)\rt\}
\\
\beta_{120}^2&=&\fr{\displaystyle \fr{2\sigma}{5x}
\lt\{\erf\lt(\fr{28}3+\fr9{x^2}\rt)-
\epa{2}{}{x}\lt(9+\fr{22x^2}3+\fr{4x^4}3\rt)\rt\}}
{\displaystyle \sqrt{\fr72}-\fr{3\alpha_s}{5x^3}\lt(-7\erf+
\epa{2x}{}{}\lt(7+\fr{22x^2}3+\fr{16x^4}3+\fr{8x^6}3\rt)\rt)}
\nonumber\\
\rule{0pt}{16mm}\trf_{121}&=&x\lt\{\erf\lt(1+\fr{77}{60x^2}-\fr9{40x^4}\rt)+
\epa{}{5}{x} \lt(\fr{14}3+\fr9{4x^2}\rt)\rt\}
\nonumber\\
\trg_{121}&=&\fr1x\lt\{\erf\lt(1-\fr7{20x^2}\rt)
+\epa{}{5}{}\lt(\fr7{2x}-\fr{7x}3-\fr{2x^3}3\rt)\rt\}
\\
\beta_{121}^2&=&\fr{\displaystyle \fr{2\sigma}{5x}
\lt\{\fr12\erf\lt(\fr{77}3-\fr9{x^2}\rt)+
\epx\lt(9-\fr{11x^2}3-\fr{2x^4}3\rt)\rt\}}
{\displaystyle \sqrt{\fr72}-\fr{3\alpha_s}{5x^3}\lt(\fr72\erf+
\xep\lt(-7+\fr{2x^2}3+\fr{8x^4}3+\fr{4x^6}3\rt)\rt)}
\nonumber\\
\rule{0pt}{16mm}\trf_{220}&=&x\lt\{\erf\lt(1+\fr1{x^2}+\fr3{4x^4}-\fr3{2x^6}\rt)+
\epx \lt(1+\fr1{2x^2}+\fr3{x^4}\rt)\rt\}
\nonumber\\
\trg_{220}&=&\fr1x\lt(1+\fr2{x^2}\rt)
\lt\{\erf\lt(1-\fr3{2x^2}\rt)+\epa{3}{}{x}\rt\}
\\
\beta_{220}^2&=&\fr{\displaystyle \fr{2\sigma}x
\lt\{\erf\lt(2+\fr3{x^2}-\fr9{x^4}\rt)+
\epa{6}{}{x}\lt(1+\fr3{x^2}\rt)\rt\}}
{\displaystyle \sqrt{\fr72}-\fr{3\alpha_s}{x^3}
\lt\{\erf\lt(-1+\fr{12}{x^2}\rt)-
\epa{2}{}{x}\lt(12+7x^2+2x^4\rt)\rt\}}
\nonumber\\
\rule{0pt}{16mm}\trf_{221}&=&x\lt\{\erf\lt(1+\fr{13}{12x^2}-\fr1{8x^4}+\fr1{x^6}\rt)+
\epx \lt(\fr23-\fr{13}{12x^2}-\fr2{x^4}\rt)\rt\}
\nonumber\\
\trg_{221}&=&\fr1x\lt\{\erf\lt(1+\fr1{4x^2}+\fr2{x^4}\rt)-
\ep\lt(\fr{2x^3}3+\fr{7x}3+\fr{19}{6x}+\fr4{x^3}\rt)\rt\}
\\
\beta_{221}^2&=&\fr{\displaystyle \fr{2\sigma}x
\lt\{\erf\lt(\fr{13}6-\fr1{2x^2}+\fr6{x^4}\rt)-
\ep\lt(\fr{2x^3}3+\fr{11x}3+\fr7{x}+\fr{12}{x^3}\rt)\rt\}}
{\displaystyle \sqrt{\fr72}-\fr{3\alpha_s}{x^3}
\lt\{-\erf\lt(\fr12+\fr8{x^2}\rt)+
\epx\lt(16+\fr{35x^2}3+6x^4+\fr{8x^6}3+\fr{4x^8}3\rt)\rt\}}
\nonumber\\
\rule{0pt}{16mm}\trf_{222}&=&x\lt\{\erf\lt(1+\fr4{3x^2}-\fr1{4x^4}-\fr1{4x^6}\rt)+
\epx \lt(1+\fr5{6x^2}+\fr1{2x^4}\rt)\rt\}
\nonumber\\
\trg_{222}&=&\fr1x\lt\{\erf\lt(1-\fr1{2x^2}-\fr1{2x^4}\rt)+
\epx\lt(\fr53+\fr1{x^2}\rt)\rt\}
\\
\beta_{222}^2&=&\fr{\displaystyle \fr{2\sigma}x
\lt\{\erf\lt(\fr83-\fr1{x^2}-\fr3{2x^4}\rt)+
\epx\lt(4+\fr3{x^2}\rt)\rt\}}
{\displaystyle \sqrt{\fr72}-\fr{3\alpha_s}{x^3}
\lt\{\erf\lt(1+\fr2{x^2}\rt)-
\epa{2}{}{x}\lt(2+\fr{7x^2}3+\fr{2x^4}3\rt)\rt\}}
\nonumber\\
\rule{0pt}{16mm}\trf_{230}&=
&x\lt\{\erf\lt(1+\fr{81}{70x^2}+\fr{33}{70x^4}+\fr{39}{28x^6}\rt)-
\epx \lt(\fr{4x^2}{35}+\fr1{35}+\fr{14}{5x^2}+\fr{39}{14x^4}\rt)\rt\}
\nonumber\\
\trg_{230}&=&
\fr1x\lt\{\erf{\textstyle \lt(1-\fr{36}{35x^2}+\fr{33}{14x^4}\rt)}-
\xep{\textstyle \lt(\fr{33}{7x^4}+\fr{26}{5x^2}+\fr{26}7+\fr{44x^2}{35}+
\fr{8x^4}{35}\rt)}\rt\}
\\
\beta_{230}^2&=&\fr{\fr{2\sigma}{7x}
\lt\{\erf\lt(\fr{81}5+\fr{66}{5x^2}+\fr{117}{2x^4}\rt)-
\ep\lt(\fr{117}{x^3}+\fr{522}{5x}+\fr{278x}5+\fr{68x^3}5
+\fr{8x^5}5\rt)\rt\}}
{\fr3{\sqrt{2}}-\fr{3\alpha_s}{7x^3}
\lt\{-6\erf\lt(\fr{12}5+\fr{11}{x^2}\rt)-
\epa{4}{}{x}\lt(33+\fr{146x^2}5+\fr{76x^4}5+\fr{32x^6}5
+\fr{12x^8}5+\fr{4x^{10}}5 \rt)\rt\}}
\nonumber\\
\rule{0pt}{16mm}\trf_{231}&=&
x\lt\{\erf\lt(1+\fr{93}{70x^2}+\fr{22}{35x^4}-\fr{13}{14x^6}\rt)+
\epa{}{7}{}
\lt(\fr{13}{x^5}-\fr2{15x^3}+\fr{23}{5x}-\fr{4x}{15}\rt)\rt\}
\nonumber\\
\trg_{231}&=&
\fr1x\lt\{\erf{\textstyle \lt(1+\fr{18}{35x^2}-\fr{11}{7x^4}\rt)}+
\epa{}{7}{}{\textstyle \lt(\fr{22}{x^3}+\fr{112}{15x}-\fr{98x}{15}-\fr{44x^3}{15}-
\fr{8x^5}{15}\rt)}\rt\}
\\
\beta_{231}^2&=&\fr{\fr{2\sigma}{7x}
\lt\{\erf\lt(\fr{93}5+\fr{88}{5x^2}-\fr{39}{x^4}\rt)-
\epa{4}{}{}\lt(\fr{39}{2x^3}+\fr{21}{5x}-\fr{107x}{30}-\fr{17x^3}{15}
-\fr{2x^5}{15}\rt)\rt\}}
{\fr3{\sqrt{2}}-\fr{12\alpha_s}{7x^3}
\lt\{\erf\lt(-\fr95+\fr{11}{x^2}\rt)+
\epa{2}{}{x}\lt(-11-\fr{83x^2}{15}-\fr{14x^4}{15}+\fr{8x^6}{15}
+\fr{2x^8}5+\fr{2x^{10}}{15}\rt)\rt\}}
\nonumber\\
\rule{0pt}{16mm}\trf_{232}&=
&x\lt\{\erf\lt(1+\fr{129}{70x^2}-\fr{121}{140x^4}+\fr{13}{56x^6}\rt)+
\epx\lt(1+\fr{149}{105x^2}-\fr{13}{28x^4}\rt)\rt\}
\nonumber\\
\trg_{232}&=&\fr1x\lt\{\erf\lt(1-\fr{36}{35x^2}+\fr{11}{28x^4}\rt)+
\ep\lt(-\fr{11}{14x^3}+\fr{23}{15x}+\fr{8x}{105}\rt)\rt\}
\\
\beta_{232}^2&=&\fr{\displaystyle \fr{2\sigma}{7x}
\lt\{\erf\lt(\fr{129}5-\fr{121}{5x^2}+\fr{39}{4x^4}\rt)+
\ep\lt(-\fr{39}{2x^3}+\fr{177}{5x}+\fr{16x}{15}\rt)\rt\}}
{\displaystyle \fr3{\sqrt{2}}-\fr{6\alpha_s}{7x^3}
\lt\{\erf\lt(\fr{36}5-\fr{11}{2x^2}\rt)+
\epx\lt(11-\fr{106x^2}{15}-\fr{52x^4}{15}-\fr{8x^6}{15}\rt)\rt\}}
\nonumber\\[5pt]
\nonumber
\end{eqnarray}

\section{String corrections}
\setcounter{equation}{0}
\renewcommand{\theequation}{B.\arabic{equation}}

The string correction potentials, $V^{\stc}_{jl\Lambda}=
\lt<\vec\Psi_{jl\Lambda}|H^{\stc}|\vec\Psi_{jl\Lambda}\rt>$, have the
following form:
\begin{eqnarray}
V^{\stc}_{110}&=&-\fr{4\sigma x}{15\beta}
\lt\{\erf\lt(1+\fr2{x^2}-\fr{11}{4x^4}\rt)
+\epx\lt(1+\fr{11}{2x^2}\rt)\rt\},
\label{scp1}
\\[10pt]
V^{\stc}_{111}&=&-\fr{\sigma x}{15\beta}
\lt\{3\erf\lt(1+\fr3{2x^2}-\fr{13}{6x^4}\rt)
+\epx\lt(5+\fr{13}{x^2}\rt)\rt\},
\\[10pt]
V^{\stc}_{120}&=&-\fr{2\sigma x}{105\beta}
\lt\{\erf\lt(\fr{28}3+\fr{67}{3x^2}-\fr3{2x^4}\rt)+\rt.
\nonumber\\
&&\lt.+\ep\lt(\fr3{x^3}-\fr{32}{3x}-16x-\fr{16x^3}3\rt)\rt\},
\\[10pt]
V^{\stc}_{121}&=&-\fr{2\sigma x}{210\beta}
\lt\{\erf\lt(\fr{77}3+\fr{139}{6x^2}+\fr3{2x^4}\rt)+\rt.
\nonumber\\
&&\lt.+\ep\lt(-\fr3{x^3}+\fr{47}{3x}-8x-\fr{8x^3}3\rt)\rt\},
\\[10pt]
V^{\stc}_{220}&=&-\fr{2\sigma x}{21\beta}
\lt\{\erf\lt(2+\fr2{x^2}-\fr3{2x^4}+\fr9{2x^6}\rt)
+\ep\lt(-\fr9{x^5}-\fr3{x^3}+\fr2x\rt)\rt\},
\\[10pt]
V^{\stc}_{221}&=&-\fr{2\sigma x}{21\beta}
\lt\{\erf\lt(\fr{13}6+\fr{59}{12x^2}+\fr3{4x^4}-\fr3{x^6}\rt)+\rt.
\nonumber\\
&&\lt.+\ep\lt(\fr6{x^5}+\fr5{2x^3}-\fr{17}{6x}-4x-\fr{4x^3}3\rt)\rt\},
\\[10pt]
V^{\stc}_{222}&=&-\fr{2\sigma x}{21\beta}
\lt\{\erf\lt(\fr83+\fr5{3x^2}+\fr3{4x^6}\rt)
+\ep\lt(-\fr3{2x^5}-\fr1{x^3}+\fr8{3x}\rt)\rt\},
\\[10pt]
V^{\stc}_{230}&=&-\fr{2\sigma x}{27\beta}
\lt\{\erf\lt(\fr{81}{35}+\fr{237}{70x^2}+\fr{87}{70x^4}-
\fr9{4x^6}\rt)+\rt.
\nonumber\\
&&\lt.+\epx\lt(\fr9{2x^4}+\fr{18}{35x^2}+\fr{131}{35}+\fr{8x^2}7
+\fr{8x^4}{35}\rt)\rt\},
\\[10pt]
V^{\stc}_{231}&=&-\fr{2\sigma x}{27\beta}
\lt\{\erf\lt(\fr{93}{35}+\fr{277}{70x^2}-\fr1{5x^4}+\fr3{2x^6}\rt)+\rt.
\nonumber\\
&&\lt.+\epx\lt(-\fr3{x^4}-\fr8{5x^2}+\fr{53}{21}+\fr{8x^2}21
+\fr{8x^4}{105}\rt)\rt\},
\\[10pt]
V^{\stc}_{232}&=&-\fr{2\sigma x}{27\beta}
\lt\{\erf\lt(\fr{129}{35}+\fr{41}{10x^2}-\fr{59}{140x^4}
-\fr3{8x^6}\rt)+\rt.
\nonumber\\
&&\lt.+\epx\lt(\fr3{4x^4}+\fr{47}{35x^2}+\fr{53}{15}\rt)\rt\}.
\end{eqnarray}

\section{Spin-orbit corrections}
\setcounter{equation}{0}
\renewcommand{\theequation}{C.\arabic{equation}}

Nonperturbative spin-orbit potentials, given as
$\VLS{jl\Lambda}=
\lt<\vec\Psi_{jl\Lambda}|H^{\LS\,(\mr{np})}|\vec\Psi_{jl\Lambda}\rt>$,
are
\begin{eqnarray}
\VLS{110}&=&\fr{2\sigma}{5\beta x}
\lt\{\erf\lt(1-\fr1{2x^2}\rt)+\epx\rt\}, \label{soc17}
\\[10pt]
\VLS{111}&=&\fr{\sigma}{5\beta x}
\lt\{\erf\lt(1+\fr1{2x^2}\rt)-\epx\rt\},
\\[10pt]
\VLS{120}&=&\fr{18\sigma}{35\beta x}
\lt\{\erf\lt(1+\fr1{2x^2}\rt)-\epx\lt(1+\fr{8x^2}9\rt)\rt\},
\\[10pt]
\VLS{121}&=&\fr{3\sigma}{35\beta x}
\lt\{\erf\lt(7-\fr3{2x^2}\rt)+\epx\lt(3-\fr{4x^2}3\rt)\rt\},
\\[10pt]
\VLS{220}&=&\fr{2\sigma}{7\beta x}\lt(1+\fr2{x^2}\rt)
\lt\{\erf\lt(1-\fr3{2x^2}\rt)+\epa{3}{}{x}\rt\},
\\[10pt]
\VLS{221}&=&\fr{2\sigma}{7\beta x}
\lt\{\erf\lt(\fr23-\fr1{2x^2}+\fr2{x^4}\rt)-
\xep\lt(\fr23+\fr5{3x^2}+\fr4{x^4}\rt)\rt\},
\\[10pt]
\VLS{222}&=&\fr{2\sigma}{7\beta x}
\lt\{\erf\lt(\fr13+\fr1{2x^2}-\fr1{2x^4}\rt)+
\epx\lt(-\fr13+\fr1{x^2}\rt)\rt\},
\\[10pt]
\VLS{230}&=&\fr{2\sigma}{63\beta x}
\lt\{{\textstyle \erf\lt(16+\fr{54}{5x^2}+\fr{21}{x^4}\rt)}-
\epa{4}{}{}{\textstyle
\lt(\fr{21}{4x^3}+\fr{31}{5x}+4x+\fr{4x^3}5\rt)}\rt\},
\\[10pt]
\VLS{231}&=&\fr{4\sigma}{9\beta x}
\lt\{\erf\lt(\fr{26}{21}+\fr3{5x^2}-\fr1{x^4}\rt)+\rt.
\nonumber\\
&&\lt.+\epa{2}{}{}\lt(\fr1{x^3}+\fr1{15x}-\fr{16x}{35}-
\fr{8x^3}{105}\rt)\rt\},
\\[10pt]
\VLS{232}&=&\fr{2\sigma}{9\beta x}
\lt\{{\textstyle \erf\lt(\fr{64}{21}-\fr{69}{35x^2}+\fr1{2x^4}\rt)}+
\ep{\textstyle \lt(-\fr1{x^3}+\fr{344}{105x}-\fr{8x}{105}\rt)}\rt\}.
\end{eqnarray}

\vspace{1cm}

Perturbative spin-dependent potentials,
$\VLSp{jl\Lambda}=
\lt<\vec\Psi_{jl\Lambda}|H^{\LS\,(\mr{p})}|\vec\Psi_{jl\Lambda}\rt>$,
read as
\begin{eqnarray}
\VLSp{110}&=&-\fr{3\alpha_s\beta}{5x^3}
\lt\{\erf+\epa{2x}{}{}\rt\},
\\[10pt]
\VLSp{111}&=&\fr35\alpha_s\beta
\lt\{\fr\erf{2x^3}-\ep\lt(2+\fr1{x^2}\rt)\rt\},
\\[10pt]
\VLSp{120}&=&\fr37\alpha_s\beta
\lt\{\fr{3\erf}{5x^3}-\epa{2}{5}{}\lt(6+\fr3{x^2}+8x^2\rt)\rt\},
\\[10pt]
\VLSp{121}&=&-\fr37\alpha_s\beta
\lt\{\fr{3\erf}{10x^3}+\epa{}{5}{}\lt(6-\fr3{x^2}+4x^2\rt)\rt\},
\\[10pt]
\VLSp{220}&=&\fr37\alpha_s\beta
\lt\{-\fr\erf{x^3}\lt(1+\fr6{x^2}\rt)+\epa{2}{}{}\lt(2+\fr5{x^2}+
\fr6{x^4}\rt)\rt\},
\\[10pt]
\VLSp{221}&=&\fr37\alpha_s\beta
\lt\{\fr{4\erf}{x^5}-\epa{4}{3}{}\lt(2+\fr5{x^4}+
\fr4{x^2}+x^2\rt)\rt\},
\\[10pt]
\VLSp{222}&=&\fr37\alpha_s\beta
\lt\{\fr\erf{x^3}\lt(1-\fr1{x^2}\rt)-\ep\lt(\fr43+\fr2{3x^2}-
\fr2{x^4}\rt)\rt\},
\\[10pt]
\VLSp{230}&=&\fr8{21}\alpha_s\beta
\lt\{\fr\erf{x^3}\lt(1+\fr9{4x^2}\rt)-\ep\lt(\fr{18}5+\fr9{2x^4}+
\fr5{x^2}+\fr{12x^2}5+\fr{8x^4}5\rt)\rt\},
\\[10pt]
\VLSp{231}&=&\fr4{21}\alpha_s\beta
\lt\{\fr\erf{x^3}\lt(1-\fr3{x^2}\rt)+\epa{2}{}{}\lt(-\fr{14}{15}+
\fr3{x^4}+\fr1{x^2}-\fr{4x^2}3-\fr{8x^4}{15}\rt)\rt\},
\\[10pt]
\VLSp{232}&=&\fr2{21}\alpha_s\beta
\lt\{\fr\erf{x^3}\lt(-4+\fr3{2x^2}\rt)+\ep\lt(\fr4{15}-
\fr3{x^4}+\fr6{x^2}-\fr{8x^2}{15}\rt)\rt\}.
\end{eqnarray}

\end{document}